\documentclass[aps]{revtex4}

\usepackage{epsfig}
\def\gsim{ \,\, \vcenter{\hbox{$\buildrel{\displaystyle >}\over\sim$}}
 \,\,}
\def\lton{ \,\, \vcenter{\hbox{$\buildrel{\displaystyle <}\over\sim$}}
 \,\,}

\newcommand{\be}{\begin{eqnarray}}
\newcommand{\ee}{\end{eqnarray}}

\newcommand{\bea}{\begin{eqnarray}}
\newcommand{\eea}{\end{eqnarray}}

\begin{document}

\thispagestyle{empty}
\title{\bf The Quark-Mass Dependence of $T_C$ in QCD:
Working up from $m=0$ or down from $m=\infty$~?}

\author{Adrian Dumitru}
\email{dumitru@th.physik.uni-frankfurt.de}
\author{Dirk R\"oder}
\email{roeder@th.physik.uni-frankfurt.de}
\author{J\"org Ruppert}
\email{ruppert@th.physik.uni-frankfurt.de}
\affiliation{Institut f\"ur Theoretische Physik,
J.~W.~Goethe Universit\"at \\
Postfach 11 19 32, D-60054 Frankfurt am Main, Germany \\
}

\begin{abstract}
We analyze the dependence of the QCD transition temperature on the
quark (or pion) mass. We find that a linear sigma model, which links
the transition to chiral symmetry restoration, predicts a much
stronger dependence of $T_c$ on $m_\pi$ than seen in present lattice data for
$m_\pi\gsim 0.4$~GeV. On the other hand, working down from
$m_\pi=\infty$, an effective Lagrangian for the Polyakov loop requires
only small explicit symmetry breaking, $b_1\sim\exp(-m_\pi)$, to
describe $T_c(m_\pi)$ in the above mass range. Physically, this is a
consequence of the flat potential (large correlation length) for the
Polyakov loop in the three-color pure gauge theory at $T_c$.
We quantitatively estimate the end point of the line of
first order deconfining phase transitions:
$m_\pi\simeq4.2\;\surd\sigma\simeq1.8$~GeV and $T_c\simeq240$~MeV 
for three flavors and three
colors.
\end{abstract}
\maketitle

\vspace*{1cm}

\section{Introduction}
Lattice QCD calculations at finite temperature and
with dynamical fermions are presently performed
for quark masses exceeding their physical values; for a recent review
see~\cite{Laermann:2003cv}. To date,
pion masses as low as $\approx400$~MeV are feasible~\cite{Karsch:2000kv},
about three times the physical pion mass. When comparing
effective theories to first-principles numerical data obtained on the
lattice it is therefore important to fix the parameters (coupling
constants, vacuum expectation values and so on) such as to match the
values of physical observables, e.g.\ of $m_\pi$, to those of the
lattice calculations. For example, the QCD equation of state in the
confined phase appears to be described reasonably well by that of a
hadron resonance gas model, {\em after extrapolating} the physical
spectrum of hadrons and resonances to that from the lattice~\cite{HRG}.
Thus, lattice data on the dependence of various observables on the
quark (or pion) mass constrain effective theories for the QCD phase
transition at finite temperature and could provide relevant
information on the driving degrees of freedom.

In this paper, we analyze the dependence of the {\em chiral} symmetry
restoration temperature on the vacuum mass of the pion using a linear
sigma model in section~\ref{sec_LinSig}.
The linear sigma model provides an effective Lagrangian
approach to low-energy QCD near the chiral limit~\cite{PW,RW}. It
incorporates the global flavor symmetry, assuming that ``color'' can
be integrated out. For example, it allows one to discuss the
order of the $N_f=2+1$ chiral phase transition as a function of the quark
masses~\cite{PW,RW,linsig3f,GGP}.

Instead of working up from zero quark mass, one could start with
the quark masses taken to infinity, that is, with a pure gauge theory.
Then, one can discuss the {\em deconfinement} transition at finite
temperature within an effective Lagrangian for the Polyakov
loop with global $Z(N_c)$
symmetry~\cite{SY,loop1,loop2,eloss,loop3,MMO,MMO2,MST} 
($N_c$ is the
number of colors). For finite pion mass, the
symmetry is broken explicitly, and the phase transition (or cross
over) temperature is shifted, relative to the pure gauge theory where
pions are infinitely heavy. In section~\ref{sec_PLoop}, we determine
the endpoint of the line of first-order transitions for three colors,
and extract the magnitude of the explicit $Z(3)$ breaking from
lattice data on $\Delta T_c$.

\section{Pion Mass and Decay Constant in Vacuum}

The Lagrangian of QCD with the quark mass matrices set to zero
is invariant under independent rotations of the $N_f$ right-handed and
left-handed quark fields. It exhibits a global $SU(N_f)_R \times
SU(N_f)_L$ symmetry, leading to $2(N_f^2-1)$ conserved currents. Those
are $N_f^2-1$ vector currents, $V_i^\mu=\bar{q}\gamma^\mu \lambda_i
q\;/2$, and $N_f^2-1$ axial currents, $A_i^\mu = \bar{q}\gamma^\mu
\gamma_5 \lambda_i q\;/2$, with $\lambda_i$ the generators of
$SU(N_f)$, normalized according to ${\rm tr}~\lambda_i
\lambda_j=2\delta_{ij}$. The $SU(N_f)_V$ subgroup of vector
transformations is preserved in the vacuum~\cite{VW}, while the
$SU(N_f)_A$ is broken spontaneously by a non-vanishing chiral
condensate $\langle\bar{q}_R q_L\rangle\neq0$, leading to
non-conservation of the axial currents.

In reality, of course, even $SU(N_f)_V$ is broken explicitly by the
non-vanishing quark mass matrix. Nevertheless, since at least $m_u$ and $m_d$
are very small in the physical limit,
the $SU(2)_V$ symmetry is almost exact in QCD. The
small explicit breaking of $SU(2)_V$ is responsible for the
non-vanishing pion mass, as given by the Gell-Mann, Oakes, Renner relation
\be \label{GOR}
m_\pi^2 = \frac{1}{f_\pi^2} \; m_q \; \langle\bar{q} q\rangle~.
\ee
We neglect isospin breaking effects here, and so assume that $m_u=m_d
\equiv m_q$.
$\langle\bar{q} q\rangle$ denotes the sum of the vacuum expectation
values of the operators $\bar{u}_R u_L$ and $\bar{d}_R d_L$, and their
complex conjugates. The proportionality constant $f_\pi$ is the pion
decay constant. It should be noted that~(\ref{GOR}) is only
valid at tree level, and that loop effects induce an implicit
dependence of both $f_\pi$ and $\langle\bar{q} q\rangle$ on $m_q$.
For small $m_q$, this dependence can be computed in chiral
perturbation theory~\cite{CPT}. For example, at next-to-leading order,
\bea
m_\pi^2 = M^2 \left[1- \frac{1}{2} \left(\frac{M}{4\pi F}\right)^2\log
\frac{\Lambda_3^2}{M^2}\right]~, \label{cPT_mpi}\\
f_\pi = F \left[ 1+ \left(\frac{M}{4\pi F}\right)^2\log
\frac{\Lambda_4^2}{M^2}\right]~, \label{cPT_fpi}
\eea
where $M$ and $F$ are the couplings of the effective theory
(equivalent to $\langle\bar{q} q\rangle$ and $m_q$), and
$\Lambda_3$ and $\Lambda_4$ are two renormalization-group invariant
scales. These relations link the behavior of $f_\pi$ to that of
$m_\pi$, the mass of a physical state. (In what follows, we use $m_\pi$
to vary the strength of explicit symmetry breaking rather than using directly
the scale dependent quark masses). 

More accurate results than eqs.~(\ref{cPT_mpi},\ref{cPT_fpi}) can
perhaps be obtained by computing quark propagators for various quark
masses on the lattice. Ref.~\cite{latt_fpi_mpi} analyzed the
propagators for gauge field
configurations generated with the standard Wilson gauge action
(``quenched QCD''), using overlap fermions with exact chiral symmetry.
They obtained a parametrisation of both $m_\pi$
and $f_\pi$ in terms of the mass $m_q$ of $u$ and $d$ quarks (see
section~2 in~\cite{latt_fpi_mpi}) which allows us to express $f_\pi$
as a function of $m_\pi$.
Their data covers an interval of
0.4~GeV$\lton m_\pi\lton1$~GeV, and 0.15~GeV$\lton \sqrt{2}
f_\pi\lton0.22$~GeV.

\section{Linear Sigma Model at Finite Temperature} \label{sec_LinSig}

In this section, we discuss chiral symmetry restoration at finite
temperature, and in particular the dependence of the symmetry
restoration temperature on the pion mass.
For simplicity, we restrict ourselves here to the two-flavor case.
Our emphasis is not on the order of the transition as the
strange quark mass is varied but rather on how the temperature at
which the transition occurs (be it either a true phase transition or
just a cross over) depends on the pion mass. 
Such dependence arises from two effects. First, of course, due to
explicit symmetry breaking occuring when $m_\pi>0$. Second, due to
the ``indirect'' dependence of spontaneous symmetry breaking, i.e.\ of
the condensate $\langle\bar{q}q\rangle$ resp.\ $f_\pi$, on the pion
mass (through pion loops, see previous section).
The tree-level potential of the linear sigma model with
$SU(2)_V\times SU(2)_A \cong O(4)$ symmetry is given by
\be
V(\sigma,\vec{\pi}) = \frac{1}{2}m^2 \phi^2 + \frac{\lambda}{4}
\phi^4- H\sigma~,
\ee
with $\phi_a=(\sigma,\vec\pi)$. For $m^2<0$ the $O(4)$ symmetry of the
vacuum state is broken spontaneously to $O(3)$, leading to a
non-vanishing scalar condensate $\langle\sigma\rangle=f_\pi$. 
The explicit symmetry breaking term $\sim H$ provides a
mass to the pions. At tree level, the masses are given by
\be
m_\sigma^2 = m^2+3\lambda\sigma^2\quad,\quad
m_\pi^2 = m^2 + \lambda\sigma^2~.
\ee

Below, we employ the Hartree approximation to investigate
the dependence of the transition temperature on the pion mass.
This approximation scheme is known to exhibit problems in the chiral
limit in that the Goldstone theorem is violated and the
phase transition is incorrectly predicted to be of first order (for
$N_f=2$). However, here we are interested only in the model with
explicit symmetry breaking, where the theory exhibits a cross
over. Specifically, we consider the region of pion masses covered by
the lattice data in~\cite{Karsch:2000kv}, $m_\pi\gsim0.4$~GeV.

In~\cite{vanHeesI} it was shown that
such truncated non-perturbative resummation schemes can be
renormalized with local counter terms obtained in the vacuum (see
also~\cite{Biancu} for $\lambda\phi^4$ theory).
These ideas were applied in~\cite{vanHeesIII} to theories with global
symmetries, and a BPHZ-like renormalization scheme was introduced for
the $O(4)$ linear sigma model in Hartree approximation
without explicit symmetry breaking. The scheme can be
straightforwardly extended to the case $H>0$, see
eqs.~(\ref{eq9}-\ref{eq11}) below. Those renormalized gap equations
coincide with
those introduced first by Lenaghan and Rischke in ref.~\cite{lenaghan}.

In this renormalization scheme a mass renormalization scale $\mu$ is
introduced and the couplings then depend on that scale (cf.\
e.g.~\cite{lenaghan}). However, choosing
\be
\mu^2=\exp\left[\frac{m_\sigma^2(\ln\,m_\sigma^2-1)-m_\pi^2(\ln\,m_\pi^2-1)}
{m_\sigma^2-m_\pi^2}\right]~,
\ee
the four-point coupling $\lambda(\mu)=\lambda_{\rm tree}$ retains its
tree-level (classical) value~\cite{lenaghan}. In other words, this
renormalization prescription evolves the renormalization scale $\mu$
in such a way as to keep $\lambda$ constant.

\begin{figure}[htp]
\centerline{\hbox{\epsfig{figure=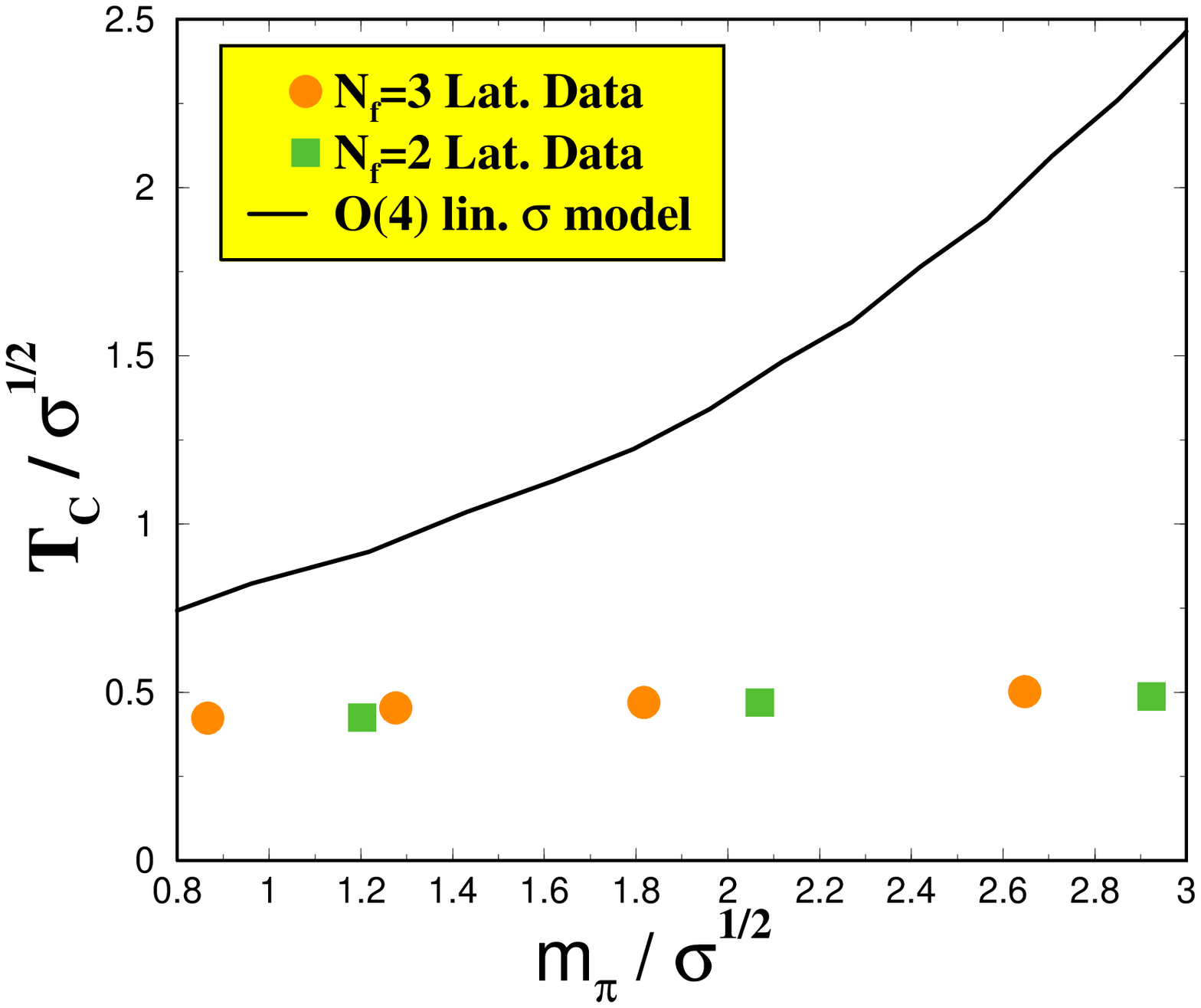,height=8cm}
\epsfig{figure=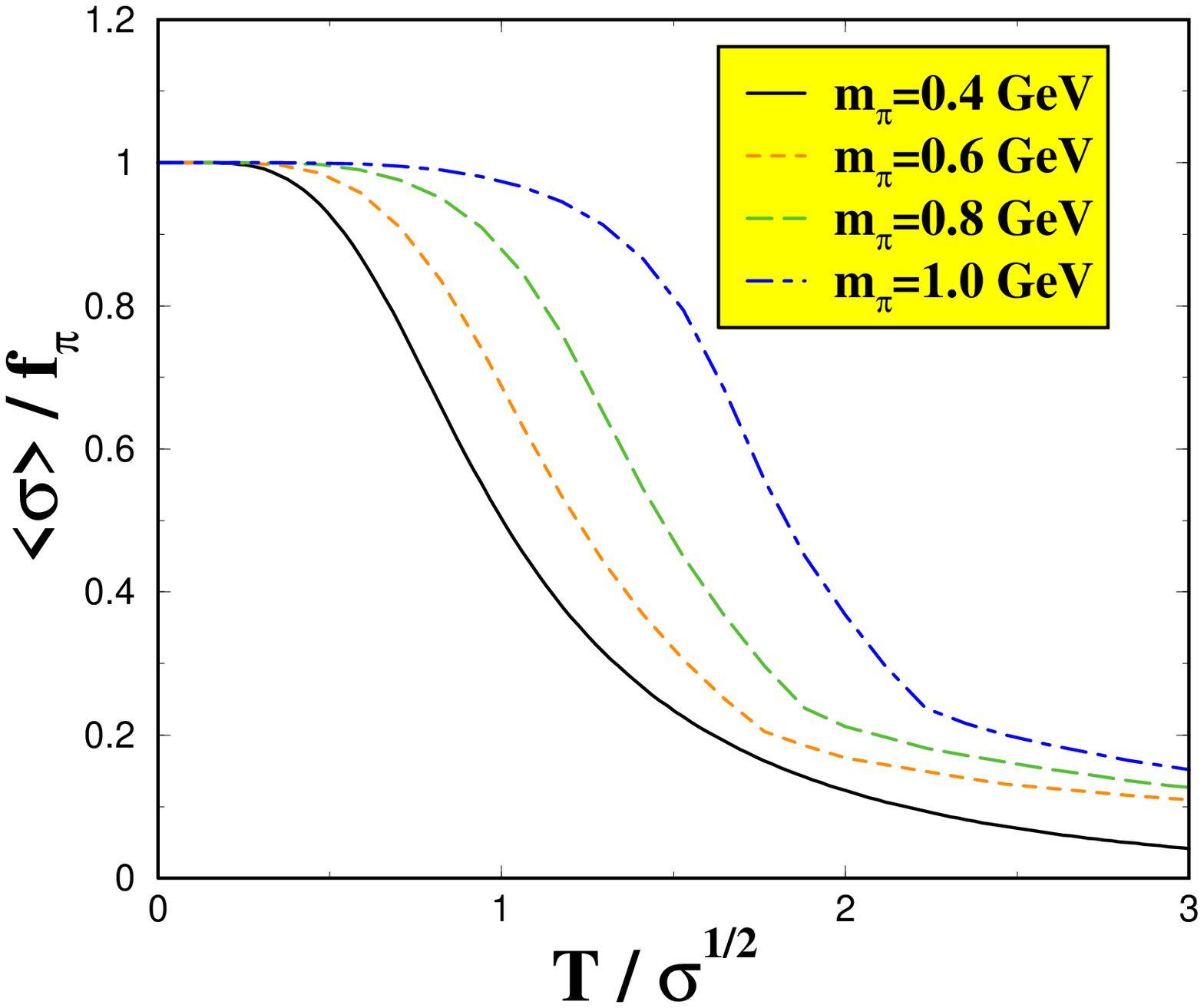,height=8cm}}}
\caption{Left: The cross over temperature $T_c$ as a function of the
(vacuum) pion mass as obtained from the linear sigma model
with $O(4)$ symmetry in comparison to lattice
data~\protect\cite{Karsch:2000kv} for two and three flavors.
The scale for both $T_c$ and $m_\pi$ is set by the zero-temperature
string tension in the pure gauge theory, $\surd\sigma\simeq0.425$~GeV.\\
Right: The scalar condensate,
$\langle\sigma\rangle$, as a function of temperature for
various pion masses.}
  \label{Fig_sigma}
\end{figure}

Explicitly, this leads to the following expressions for the
couplings~\cite{lenaghan}:
\be
\lambda = \frac{1}{2} \frac{m_\sigma^2-m_\pi^2}{f_\pi^2}\quad,\quad
 H=f_\pi\left(m_\sigma^2-2\lambda f_\pi^2 \right)\quad,\quad
m^2 = -\frac{1}{2} \left( m_\sigma^2-3 m_\pi^2\right)
-6\lambda Q_\mu(m_\pi)\quad,
\ee
where
\be
Q_\mu(M)\equiv \frac{1}{(4\pi)^2}\left[M^2\ln\frac{M^2}{\mu^2}-M^2+\mu^2\right].
\ee
These equations determine the couplings in vacuum in terms of $m_\pi$,
$f_\pi$ and $m_\sigma$. The dependence of $f_\pi$ on $m_\pi$ is
taken from the data of ref.~\cite{latt_fpi_mpi} (cf.\ their figs.~1,
2 and the corresponding fits therein),  as mentioned above.
Roughly, for $m_\pi:0.4~{\rm GeV}\to 1$~GeV, $f_\pi$
increases by about 50~\%, leading to an increase of the explicit
symmetry breaking term $H$ by a factor of 10.
We also require the dependence of $m_\sigma$
on $m_\pi$, which we take from a recent computation with standard
Wilson fermions~\cite{latt_sigma}. Those authors
find that $m_\sigma$ is essentially a linear function of
$m_\pi^2$. We checked how our results in Fig.~\ref{Fig_sigma} depend
on this assumption by using, alternatively, a linear dependence
$m_\sigma= m_\pi+{\rm const.}$, with
$m_\sigma=0.6$~GeV for $m_\pi=0.14$~GeV. We found essentially the same
dependence of $T_c$ on $m_\pi$.

At finite temperature, we use the 
effective potential for composite operators~\cite{CJT} to determine
the masses and the scalar condensate in the Hartree approximation.
We follow the derivation outlined in~\cite{lenaghan,RRR}. The
resulting gap equations are 
\bea
 H &=& \langle \sigma \rangle \left[m_\sigma^2-2\lambda \langle \sigma
       \rangle^2 \right] \quad, \label{eq9}\\
 m_\sigma^2 &=& m^2+3 \lambda \left\{\langle \sigma \rangle^2 + 
[Q_T(m_\sigma)+Q_\mu(m_\sigma)]+
[Q_T(m_\pi)+Q_\mu(m_\pi)]\right\} \quad,  \label{eq10}\\
 m_\pi^2 &=& m^2+ \lambda \left\{\langle \sigma \rangle^2 + 
[Q_T(m_\sigma)+Q_\mu(m_\sigma)]
+5[Q_T(m_\pi)+Q_\mu(m_\pi)]\right\} \quad, \label{eq11}
\eea
where the finite temperature contribution of the tadpole-diagram
is given by
\be
 Q_T(m) \equiv \int 
 \frac{d^3k}{(2 \pi)^3} \, \frac{1}{\epsilon_k(m)}
 \frac{1}{\exp[\epsilon_k(m)/T]-1}\quad,\quad
\epsilon_k(m)\equiv \sqrt{\vec{k}^2+m^2}~.
\ee
Here, $m_\sigma$, $m_\pi,$ and $\langle \sigma \rangle$ denote the
effective masses and the scalar condensate at finite temperature,
respectively. 
The self-consistent solution of the above
gap equations for a given vacuum pion mass determines
the temperature dependence of the scalar condensate as the order
parameter of chiral symmetry restoration. For explicitly broken chiral
symmetry, $H>0$, the transition in this approach is a cross over. We
define the cross over temperature $T_c$ by the peak of
$\partial \langle \sigma \rangle / \partial T$.

The dependence of $T_c$ on $m_\pi$ is depicted in Fig.~\ref{Fig_sigma}
(left), where we have also shown lattice results obtained with two and
three degenerate quark flavors, respectively~\cite{Karsch:2000kv} (the
$N_f=2$ data with standard action, the $N_f=3$ data with improved
p4-action). Driven by the increase of both $f_\pi$ and $H$ with
$m_\pi$, the linear sigma model predicts a rather rapid rise of $T_c$
with the pion mass, as compared to the data which is nearly flat on the
scale of the figure. While lattice data indicate a rather weak
dependence of $T_c$ on the quark mass (see also ref.~\cite{latt2}),
models with spontaneous symmetry breaking in the vacuum
naturally predict a rather steep rise of $T_c$
with the VEV $\langle\sigma\rangle_{\rm vac}=f_\pi$, which itself
increases with the quark (or pion) mass. Our findings here are in
qualitative agreement with those from ref.~\cite{BJW} who employed
nonperturbative flow equations to compute the effective potential
for two flavor QCD within the linear sigma model. They also find a
steeper slope of $T_c(m_\pi)$ than indicated by the lattice, even though their
analysis appears to predict a somewhat weaker increase of $f_\pi$
with $m_\pi$ than the data of~\cite{latt_fpi_mpi}, which we employ here.

Fig.~\ref{Fig_sigma} also shows the temperature dependence of the
$\sigma$ condensate (right). With $m_\sigma$ a linear function of
$m_\pi^2$~\cite{latt_sigma}, the width of the cross over is
approximately independent of 
the pion mass for 0.4~GeV$\lton m_\pi\lton1$~GeV, while we found 
considerable broadening when $m_\sigma$ is linear in $m_\pi$ (not shown).
The chiral susceptibility $\partial\langle\sigma\rangle/\partial T$ at
its maximum is $\approx 0.25$, i.e.\ the cross over is in fact
quite broad for the range of $m_\pi$ considered. Since this is
at variance with lattice data on QCD thermodynamics (pressure
and energy density as functions of temperature, see e.g.\ the review
in~\cite{Laermann:2003cv}), one might argue that the cross over is in
fact not driven by the order parameter field but by heavier degrees of
freedom~\cite{HRG,CPT}. Such degrees of freedom could reduce
the pion-mass dependence of the transition substantially: using
three-loop chiral perturbation theory (i.e.\ the non-linear model),
Gerber and Leutwyler find~\cite{CPT} that $T_c$ increases rapidly from
$\approx 190$~MeV in the chiral limit
(using their set of couplings) to $\approx 240$~MeV for
physical pion mass. However, when heavy states are included (in the
dilute gas approximation), then $T_c$ increases less rapidly, from
$\approx 170$~MeV in the chiral limit to $\approx 190$~MeV for
physical pion mass. 

Hence, perhaps the transition is not primarily driven by an order
parameter field and infrared dynamics.
Another possible approach is discussed in the next section.

\section{Effective Lagrangian for the Polyakov loop at Finite
Temperature} \label{sec_PLoop}

In the limit $m_q\to\infty$ the quarks decouple and drop out of the theory.
The (gauge-invariant) order parameter for the deconfining phase
transition in such a pure gauge theory with $N_c$ colors is the Polyakov loop
\be \label{PloopDef}
\ell = \frac{1}{N_c} \; {\rm tr}\; {\cal P} 
\exp \left( i g \int^{1/T}_0 A_0(\vec{x},\tau) \, 
d\tau \right)~.
\ee
$A_0$ denotes the temporal component of the gauge field in the
fundamental representation, $g$ is the gauge
coupling, and path ordering is with respect to imaginary time $\tau$.
Its expectation value, $\ell_0(T)$, vanishes when $T< T_c$, and
is nonzero above $T_c$.  Indeed, by asymptotic freedom,
$\ell_0 \rightarrow 1$ as $T\rightarrow \infty$.
The simplest guess for a potential for the Polyakov loop is:
\begin{equation}
V(\ell) = - \frac{b_2}{2} |\ell|^2 
+ \frac{1}{4} \left( |\ell|^2 \right)^2 \; \quad\quad(N_c=2)~.
\label{e3}
\end{equation}
The  Polyakov loop model~\cite{loop1,loop2,eloss,loop3} is a mean field
theory for $\ell$. In a mean field analysis
all coupling constants are taken as constant with temperature, except
for the mass term, $\sim b_2 |\ell|^2$.  
About the transition, condensation
of $\ell$ is driven by changing the sign of the two-point coupling:
$b_2 > 0$ above $T_c$ ($b_2(T)\to 1$ for $T\to\infty$),
and $<0$ below $T_c$.

For two colors, (\ref{e3}) is a mean field theory for a second order
deconfining transition \cite{second}.
The $\ell$ field is real, and so the potential defines a mass:
$(m_\ell/T)^2 = (1/Z_s) \partial^2 V/\partial\ell^2$,
where $Z_s$ is the wave function normalization constant for
$\ell$~\cite{wirstam}. The mass is measured
from the two point function of Polyakov loops in coordinate space, 
$\propto (1/r)\exp(- m_\ell\; r)$ as $r\rightarrow \infty$.

For three colors, $\ell$ is a complex valued field,
and a term cubic in $\ell$ appears in $V(\ell)$,
\begin{equation}
V(\ell) = - \frac{b_2}{2} |\ell|^2 
- \frac{b_3}{3}\frac{\ell^3 + \ell^*\!^3}{2}
+ \frac{1}{4} \left( |\ell|^2 \right)^2 \; \quad\quad(N_c=3)~.
\label{e3_N=3}
\end{equation}
At very high
temperature, the favored vacuum is perturbative, with $\ell_0 \approx 1$,
times $Z(3)$ rotations.  We then choose $b_3>0$ so that in the $Z(3)$ model,
there is always one vacuum with a real, positive expectation value for
$\ell_0$.
This produces a first order deconfining transition, where $\ell_0$ jumps
from $0$ at $T_c^-$ to $\ell_c = 2b_3/3$ at $T_c^+$~\cite{loop2,eloss};
$T_c$ is given by $b_2(T_c) = -2b_3^2/9$.
The $\ell$ field has two masses, from its 
real ($m_\ell$) and imaginary ($\widetilde{m}_\ell$) parts.
At $T_c^+$, $\sqrt{Z_s} m_\ell/T = \ell_c$.
The mass for the imaginary part of $\ell$ is
$\sqrt{Z_s} \widetilde{m}_\ell(T)/ T \propto \sqrt{b_3 \ell}$;
at $T_c^+$, $\widetilde{m}_\ell/m_\ell = 3$, twice the value expected
from a perturbative analysis of the loop-loop correlation function,
obtained by expanding $\ell$ from eq.~(\ref{PloopDef}) to order
$A_0^3$~\cite{2point}. This mass ratio receives corrections if
five-point and six-point couplings are included in the effective
Lagrangian~\cite{2point} but those are not crucial for the present
discussion. We note that, in principle, all of the above coupling
constants could be determined on the lattice.
The lattice regularization requires non-perturbative
renormalization of the Polyakov loop in order to define the proper
continuum limit of $\ell$~\cite{PLren,flatrenPLpot}.

Within the above mean-field theory, dynamical quarks act like a
``background magnetic field'' which breaks the $Z(3)$ symmetry
explicitly, and a term linear in $\ell$
also appears in $V(\ell)$~\cite{banks,Green:1983sd,MeisOg95,Alexandrou:1998wv}:
\begin{equation}
V(\ell) = - b_1 \frac{\ell + \ell^*}{2} 
- \frac{b_2}{2} |\ell|^2 
- \frac{b_3}{3}\frac{\ell^3 + \ell^*\!^3}{2}
+ \frac{1}{4} \left( |\ell|^2 \right)^2 \; \quad\quad(N_c=3,~m_\pi<\infty)~.
\label{e3_Q}
\end{equation}
Hence, as $m_\pi$ decreases from infinity, $b_1(m_\pi)$ turns on. The
normalization of $b_2(T)$ for $T\to\infty$ is such that
$\ell_0\to 1$, i.e.\ $b_2(T=\infty)=1-b_1-b_3$. 

We first consider the case where $b_1$ is very small, and take
the term linear in $\ell$ as a perturbation; then
the weakly first-order phase transition of the
pure gauge theory persists (in what follows, the critical temperature
in the pure gauge theory with $b_1=0$ will be denoted $T_c^*$). 
The critical temperature is determined from
\be \label{b2Tc}
b_2(T_c) = -\frac{2}{9} b_3^2 \left(1+\frac{27}{2}\frac{b_1}{b_3^3}
\right) + {\cal O}(b_1^2)~.
\ee
The order parameter jumps at $T_c$,
from 
\be \label{ellTc-}
\ell_0(T_c^-) = \frac{9}{2} \frac{b_1}{b_3^2} + {\cal O}(b_1^2)~,
\ee
to
\be \label{ellTc+}
\ell_0(T_c^+) = \frac{2}{3}b_3 - \frac{9}{2} \frac{b_1}{b_3^2}
                   + {\cal O}(b_1^2)~. 
\ee
Note that numerically $\ell_0(T_c^-)$ could be much larger than $b_1$
if the phase transition in the pure gauge theory is weak and so the
correlation length $\xi=1/m_\ell$ near $T_c$ is large (i.e.\ if
$b_3$ is small), as indeed appears to be the case for $N_c=3$
colors~\cite{mell}. In other words, it could be that on the lattice
$\ell$ quickly develops a non-vanishing expectation value at $T_c^-$
already for rather large quark (or pion) masses, but this does not
automatically imply a large explicit symmetry breaking (see also
Fig.~\ref{Fig_ell} below).

{}From eq.~(\ref{b2Tc}) we can estimate the shift of $T_c$ induced by
letting $m_\pi<\infty$.
Writing the argument of $b_2$ in that equation as
$T_c^* + \Delta T_c$ and expanding to first order in $\Delta T_c$ we
obtain
\be \label{DelTc}
\frac{\Delta T_c}{T_c^*} = -3\; \frac{b_1}{b_3}~\left(
T\frac{\partial b_2}{\partial T}\right)^{-1}_{T=T_c^*} + {\cal O}(b_1^2)= 
-\frac{2}{3}~\ell_0(T_c^-)~b_3\left(
T\frac{\partial b_2}{\partial T}\right)^{-1}_{T=T_c^*}+ {\cal O}(b_1^2)~.
\ee
The shift in $T_c$ with decreasing pion mass is proportional to
the expectation value of the Polyakov loop just below $T_c$; all other
factors on the right-hand side of eq.~(\ref{DelTc}) do not depend on
$b_1$ or $m_\pi$.
Numerical values for $b_3$ and for $b_2(T)$ were obtained
in~\cite{loop2,loop3,Ove} by fitting the effective potential~(\ref{e3_N=3})
to the pressure and energy density of the pure gauge theory with three
colors; those are $b_3\approx0.9$ and $b_3^2/(T_c^* \partial
b_2(T_c^*)/\partial T)\approx 1$, to within 10\%. We therefore expect
that numerically $\Delta T_c/T_c^*$ is roughly equal to $\ell_0(T_c^-)$.

Eqs.~(\ref{ellTc-},\ref{ellTc+}) seem to indicate that
the discontinuity of $\ell_0$ at $T_c$ vanishes, i.e.\ that the phase
transition turns into a cross over, at a pion mass such that 
$b_1(m_\pi)=2b_3^3/27$. However, we can not really extend our ${\cal
O}(b_1)$ estimates to the end point of the line of first-order
transitions because it applies, near $T_c$, only if $-4b_2(T_c) \ll
b_3^2$, which translates into $b_1 \ll b_3^3/108$, see eq.~(\ref{b2Tc}).
To find the endpoint of the line of first-order transitions we
solve numerically for the global minimum of~(\ref{e3_Q}) as a function
of $b_2$, for given $b_1$, see Fig.~\ref{Fig_ell} (left). The
numerical solution is ``exact'' and does not assume
that $b_1$ is small. We employ $b_3=0.9$ to properly account for the
small latent heat of the pure gauge
theory~\cite{loop2,eloss,loop3,2point,Ove}. 
Also, for $b_1=0$, this $b_3$ corresponds to $\ell_c=0.6$, which is close
to the expectation value of the renormalized (fundamental) loop for
the $N_c=3$ pure gauge theory~\cite{PLren,flatrenPLpot}.

Clearly, for very small $b_1$ the order parameter $\ell_0$ jumps at
some $b_2^c\equiv b_2(T_c)$, i.e.\ the first-order phase
transition persists. (The abscissa is normalized by
$|b_2(T_c^*)| = 2b_3^2/9$.)
We find that the discontinuity vanishes at $b_1^c=0.026(1)$, so
there is no true phase transition for $b_1 > b_1^c$. Nevertheless,
we define $b_2^c$ even in the cross over regime via
the peak of $\partial\ell_0(b_2)/\partial b_2$.
The shift of $b_2^c$ with increasing $b_1$ can now be converted
into the shift of $T_c$ itself by expanding about $T_c^*$:
\be \label{DelTc2}
\frac{\Delta T_c}{T_c^*} = {\Delta b_2^c}~\left(
T\frac{\partial b_2}{\partial T}\right)^{-1}_{T=T_c^*}~,
\ee
as already discussed above. We also note that from Fig.~\ref{Fig_ell}
(left), the susceptibility for the Polyakov loop at its maximum is
$\partial \ell_0/\partial b_2\simeq 3.5$, 2, 1.5 for $b_1=0.06$, 0.1,
and 0.126, respectively. That is, the cross over is rather sharp for
the values of $b_1$ shown in the figure.

Explicit breaking of the Z(3) symmetry of the gauge theory
has previously been studied
in~\cite{Green:1983sd,MeisOg95,Alexandrou:1998wv}, and has been 
identified as the essential factor in determining the endpoint of
deconfining phase transitions. Moreover, while the term $\sim b_1$
quickly washes out the transition, those studies showed that along the
line of first order transitions the shift of $T_c$ (or, alternatively, of
the critical coupling $\beta_c$) is moderate, which agrees with
our findings. However, the numerical values for the critical ``external field"
at the endpoint obtained in~\cite{MeisOg95,Alexandrou:1998wv} from actual
Monte-Carlo simulations can not be compared directly to our estimate for
$b_1^c$ because we work here with the renormalized (continuum-limit) loop,
not the bare loop.

\begin{figure}[htp]
\centerline{\hbox{\epsfig{figure=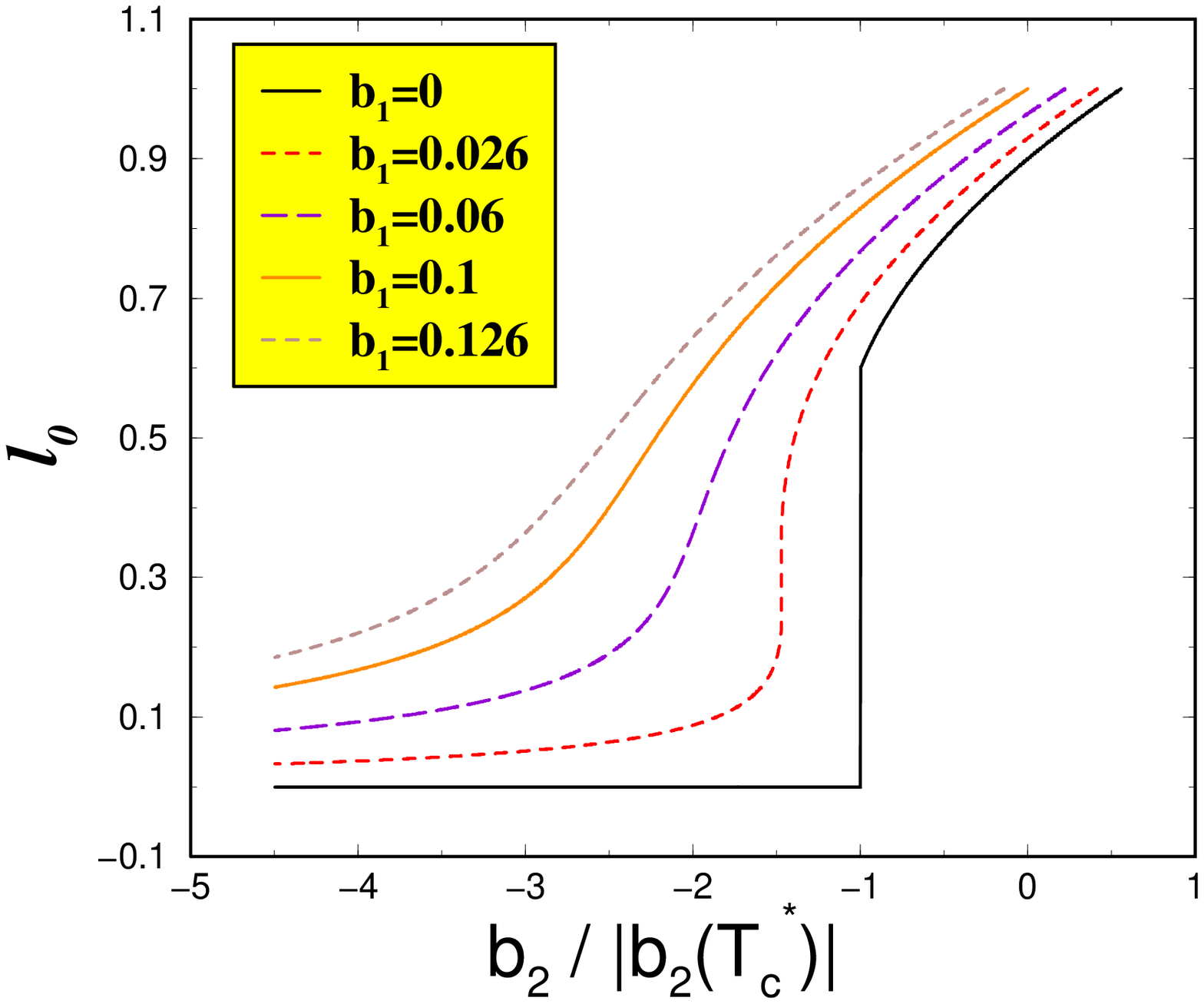,height=8cm}
\epsfig{figure=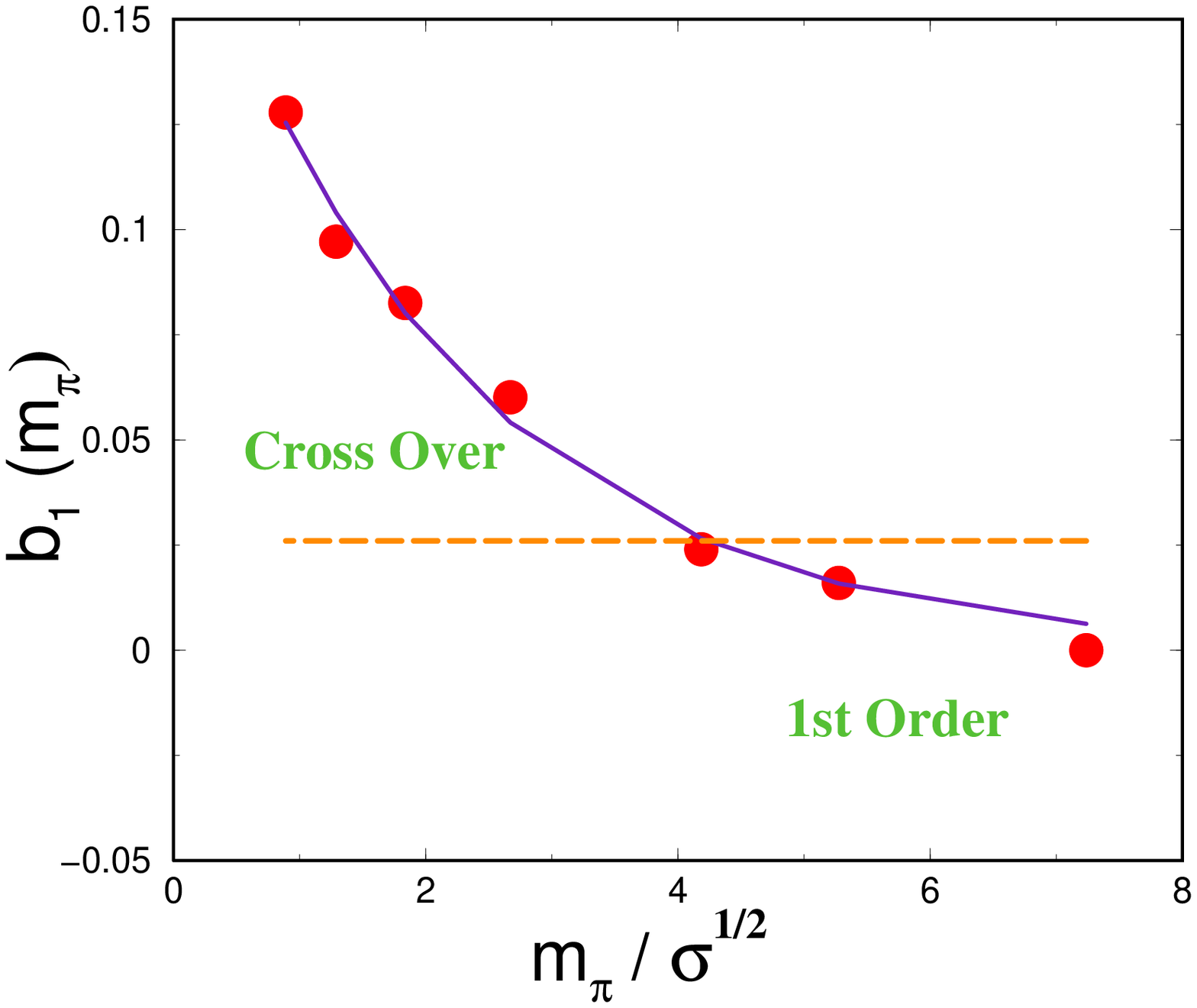,height=8cm}}}
\caption{Left: The expectation value for the Polyakov loop,
$\ell_0(b_2(T))$, for various values of the explicit
symmetry breaking coupling, $b_1$. All curves terminate at
$\ell_0=1\Leftrightarrow T=\infty$.\\
Right: $b_1$ as a function of $m_\pi$, obtained by matching to three flavor
lattice data for $T_c(m_\pi)$. The solid line corresponds to an exponential
increase of $b_1$ with decreasing $m_\pi$, see text.
The broken horizontal line displays the
endpoint of the line of first-order phase transitions in terms of
$b_1$; the intersection with the $b_1(m_\pi)$ curve then gives the
corresponding pion mass.}
  \label{Fig_ell}
\end{figure}
Ref.~\cite{Karsch:2000kv} studied finite-temperature
QCD with $N_f=3$ flavors and various quark masses on the lattice (with
improved p4-action), and determined the critical (or cross over)
temperature as a
function of the pion mass. Using eq.~(\ref{DelTc2}) we can
match our $\Delta T_c/T_c^*$ to the data from~\cite{Karsch:2000kv} to
determine $b_1(m_\pi)$. In other words, we extract the
function $b_1(m_\pi)$ required to match the effective
Lagrangian~(\ref{e3_Q}) to $T_c(m_\pi)$ found on the lattice.
The result is shown in Fig.~\ref{Fig_ell} on the right. (Again, the pion mass
is normalized to the zero-temperature string tension in the pure gauge
theory, $\surd\sigma\simeq0.425$~GeV.)

Evidently, the $\approx33\%$ reduction of $T_c$ from $m_\pi=\infty$
(pure gauge theory) to $m_\pi/\surd\sigma\approx 1$ requires only
small explicit breaking of the $Z(3)$ symmetry for the Polyakov loop
$\ell$: we find that $b_1<0.15$ even for the smallest pion
masses available on the lattice. This is due to the rather weak first-order
phase transition of the pure gauge theory with $N_c=3$ colors,
reflected by the strong dip of the string tension 
in the confined phase near $T_c^-$ and of the Polyakov loop screening
mass $m_\ell$ in the deconfined phase near $T_c^+$~\cite{mell}; cf.\
also the discussion in~\cite{loop1,loop2,eloss,2point}.

Moreover, $b_1(m_\pi)$ appears to follow the
expected behavior $\sim\exp(-m_\pi)$. The exponential fit shown by the
solid line corresponds to $b_1(m_\pi) = a \exp(-b \; m_\pi/\surd\sigma)$,
with $a=0.19$ and $b=0.47$~. Surprisingly, by naive extrapolation
one obtains a pretty small
explicit symmetry breaking even in the chiral limit,
$b_1\approx 0.2$.

Finally, the endpoint of the line of
first-order transitions at $b_1^c=0.026$ (indicated by the dashed
horizontal line) intersects the curve $b_1(m_\pi)$ at
$m_\pi/\surd\sigma\approx 4.2$. For heavier pions the theory exhibits a
first-order deconfining phase transition, which then turns into a
cross over for $m_\pi\lton4.2\;\surd\sigma\approx1.8$~GeV. 
According to our estimate, the
endpoint of the line of first order transitions occurs at $\Delta
T_c/T_c^*\approx 12\% $, which is slightly less than a previous
(qualitative) estimate of 26\% from ref.~\cite{MMO2}.

Of course, so far our analysis is restricted to pion masses
$m_\pi/\surd\sigma\gsim1$. On the other hand,
one might cross a chiral critical point for some pion
mass $m_\pi/\surd\sigma<1$~\cite{GGP}. Attempting a fit
with the model~(\ref{e3_Q}) beyond that point would then lead to 
deviations from $b_1\sim\exp(-m_\pi)$.

\section{Discussion}
Three-color QCD exhibits a (weakly) first-order deconfining phase
transition at a temperature $T_c/\surd\sigma\approx0.63$ in the limit
of infinitely heavy quarks ($\surd\sigma\approx0.425$~GeV denotes the
string tension at $T=0$ in this theory). Near $T_c$,
the screening mass for the fundamental Polyakov loop $\ell$ drops
substantially~\cite{mell}, and so one might hope to capture the
physics of the phase transition using some effective Lagrangian for
$\ell$~\cite{SY,loop1,loop2,eloss,loop3,MMO,MMO2,MST}.

\begin{figure}[htp]
\vspace*{-1cm}
\centerline{\hbox{\epsfig{figure=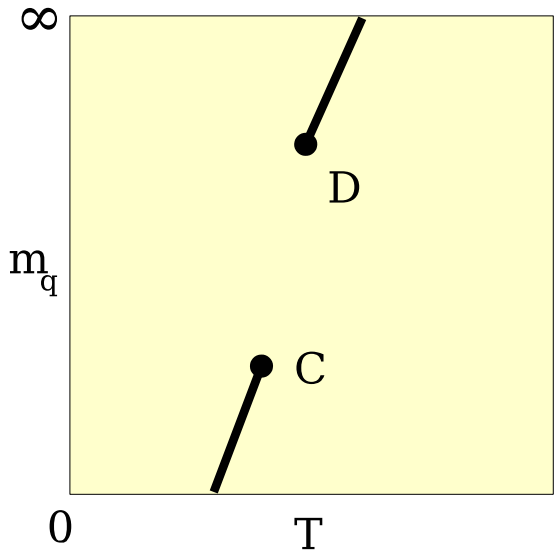,height=12cm}}}
\vspace*{-1.7cm}
\caption{Schematic phase diagram in the temperature vs.\ quark mass
plane~\protect\cite{GGP}. ${C}$ is the chiral critical point, 
${D}$ the deconfining critical point.}
  \label{Fig_ggp}
\end{figure}
For finite quark masses, a term linear in $\ell$ appears which
breaks the $Z(3)$ center-symmetry explicitly. This reduces the
deconfinement temperature, with $\Delta T_c/T_c^*$ on the order of
the expectation value of the Polyakov loop at $T_c^-$,
cf.\ eq.~(\ref{DelTc}).

At some point then,
the line of first-order deconfinement phase transitions
ends~\cite{GGP,Green:1983sd,MeisOg95,Alexandrou:1998wv,Brown:1990ev}, see
Fig.~\ref{Fig_ggp}. We have provided 
a quantitative estimate of this point, $m_\pi \simeq 4.2\;\surd\sigma
\simeq1.8$~GeV and $T_c\simeq240$~MeV for
$N_f=3$ degenerate flavors, by matching our effective
Lagrangian for the Polyakov loop to lattice data on
$T_c(m_\pi)$~\cite{Karsch:2000kv}. Assuming that $b_1\propto
N_f$~\cite{Alexandrou:1998wv} shifts ``D'' to $m_\pi \simeq 1.4$~GeV
for $N_f=2$ and to 0.8~GeV for $N_f=1$.

Going to even smaller quark (or pion) masses leaves a cross over
between the low-temperature and high-temperature regimes of QCD.
The dependence of the cross over temperature $T_c$
on the pion mass appears to be well described by a small explicit
breaking of the $Z(3)$ center symmetry, $b_1\sim\exp(-m_\pi)$, down to
$m_\pi/\surd\sigma\simeq1$, which is the smallest pion mass covered by
the lattice data of ref.~\cite{Karsch:2000kv}. On the other hand, a
linear sigma model leads to a stronger dependence of $T_c$ on
$m_\pi$ than seen in the data. 

In turn, in the chiral limit, and for $N_f=3$ flavors, one
expects a first-order {\em chiral} phase
transition~\cite{PW,GGP,Brown:1990ev}. The linear sigma model should then
be an appropriate effective Lagrangian
for low-energy QCD~\cite{PW,RW,linsig3f,GGP}. The first-order chiral
phase transition ends in a critical point ``C'' if either the mass of
the strange quark or those of all three quark flavors increase. Given
that the explicit symmetry breaking term for the Polyakov loop remains
rather small when extrapolated to $m_\pi\to0$, that is $b_1\to0.2$,
we speculate that ``C'' might be rather close to the chiral
limit. Indeed, recent lattice estimates for $N_f=3$ place ``C'' at $m_\pi\simeq
290$~MeV~\cite{lattC} for standard staggered fermion action and
$N_t=4$ lattices; improved (p4) actions predict values as low as
$m_\pi\simeq67$~MeV~\cite{Karsch:2003va}.

Of course, the question arises why, for pion masses down to
$\simeq400$~MeV, the QCD cross over is described rather naturally by
a slight ``perturbation'' of the $m_\pi=\infty$ limit, in the form of
an explicit breaking of the global $Z(N_c)$ symmetry for the Polyakov
loop.
Physically, the reason is the flatness of the potential for $\ell$ in
the pure gauge theory at $T_c$, see e.g.\ the figures
in~\cite{loop2,Ove}, which causes the sharp drop of the screening mass
for $\ell$ near $T_c^+$~\cite{mell}. 
This is natural, given that finite-temperature expectation values of
Polyakov loops at $N_c=3$ are close to those at
$N_c=\infty$~\cite{flatrenPLpot}, where the
potential at $T_c$ becomes entirely flat~\cite{GW,flatrenPLpot}.
Hence, a rather small ``tilt'' of the
potential (due to explicit symmetry breaking) quickly washes out the
deconfining
phase transition of the pure gauge theory, and causes a significant
shift $\Delta T_c$ of the cross over temperature already for small $b_1$.
If so, then for $N_c\to\infty$, at the Gross-Witten
point, the endpoint ``D'' should be located at
$b_1=0$; the discontinuity for the Polyakov loop at $T_c$,
which in a mean-field model for the pure gauge theory  is 1/2 at
$N_c=\infty$~\cite{flatrenPLpot,GW} then vanishes for arbitrarily
small explicit symmetry breaking. This has previously been noted by
Green and Karsch~\cite{Green:1983sd} within a mean-field model. If
confirmed by lattice Monte-Carlo studies, we might improve our
understanding of the degrees of freedom driving the QCD cross over for
pion masses above the chiral critical point ``C''.\\[.5cm]
{\bf Acknowledgement:}\\
We thank H.~van Hees, F.~Karsch, J.~Lenaghan, A.~Peshier, R.~Pisarski, and
D.H.~Rischke for useful
comments. A.D.\ acknowledges support from BMBF and GSI and
J.R.\ from the Studienstiftung des
deutschen Volkes (German National Merit Foundation).

\end{document}